\RequirePackage{fix-cm}

 \documentclass[showkeys,preprint,nofootinbib]{revtex4}          % twocolumn%
\usepackage{graphicx}
\usepackage{amssymb}
\usepackage{amsbsy}
 \usepackage{amsmath}
\usepackage{subcaption}
\usepackage{color}
\usepackage{enumitem}
\usepackage{easybmat}
\usepackage{xypic}
\usepackage{comment}
\usepackage{multirow}

\newcommand{\rd}{\mathrm{d}}

\usepackage{MnSymbol,wasysym}

\begin{document}

\title[]{Generalized and graded geometry for mechanics: \\a comprehensive introduction}

\author{Vladimir Salnikov} 
\email{vladimir.salnikov@univ-lr.fr}
\affiliation{
LaSIE  -- CNRS \& La Rochelle University,
UMR CNRS 7356, Av. Michel Cr\'epeau, 17042 La Rochelle Cedex 1, France}

\author{Aziz Hamdouni} 
\email{aziz.hamdouni@univ-lr.fr}
\affiliation{
LaSIE  -- La Rochelle University,
UMR CNRS 7356, Av. Michel Cr\'epeau, 17042 La Rochelle Cedex 1, France}

\author{Daria Loziienko} 
\email{daria.loziienko1@univ-lr.fr}
\affiliation{
LaSIE  -- La Rochelle University,
UMR CNRS 7356, Av. Michel Cr\'epeau, 17042 La Rochelle Cedex 1, France}

\begin{abstract}

In this paper we make an overview of results relating the recent ``discoveries'' in differential geometry, such as \emph{higher structures} and \emph{differential graded manifolds} with some natural problems coming from mechanics.  We explain that a lot of classical differential geometric constructions in the context can be conveniently described using the language of $Q$-structures, and thus $Q$-structure preserving integrators are potentially of great use in mechanics. We give some hints how the latter can be constructed, and formulate some open problems. 

Since the text is intended both to mathematics and mechanics communities, we tried to make it accessible to non-geometers as well.



\end{abstract}

\maketitle

\section{Introduction}

This paper is an attempt to put together several results of the authors, united under the same manifest: ``geometry encodes the physics of the system for reliable modelling and efficient computer simulations''. By this we mean that it is useful and sometimes even necessary to take into account the internal structure of the equations governing the dynamics of a mechanical system. This internal structure is often encoded using the concepts from differential geometry, preserving them in computations typically results in better quality, i.e. more reliable  simulation results. 

Disclaimer: \textit{This text is a recollection of several talks we have delivered at different editions of the CITV meeting\footnote{Colloque International de Th\'eories Variationnelles,  Souriau Colloquium.} we will stick to Souriau's style -- paying attention to conceptual ideas rather than to technicalities. We will however do our best to honestly introduce all the necessary notions, addressing both the specialists in geometry and mechanics.} In the main text of this paper we will assume that the reader is familiar with basic notions of classical differential geometry, if this is not the case we have added the appendix \ref{sec:geom} to make the presentation self-contained.

The following table shows the most commonly encountered properties of  mechanical systems and the geometric structures that are appropriate to describe them. 
\begin{center}
 \begin{tabular}{|c|c|c|}
 \hline
 & Mechanical property& Geometric description \\
 \hline
\multirow{6}{*}{ 
  \begin{minipage}{0.24\linewidth} \begin{center} \vspace{-1em}
   classical  \\ classical \\ mechanics \\ (ODE)    \end{center} \vspace{-2em}
  \end{minipage} }
    & conservation of energy	&   Poisson / \textbf{symplectic}  \\ \cline{2-3}
    &\multirow{2}{*}{ symmetries }& Lie groups/algebras, \\
    & & \textbf{Cartan moving frames}\\ \cline{2-3}
    & dissipation / interaction   & \multirow{2}{*}{\textbf{(almost)} Dirac} \\
   & power balance; constraints &  \\   \cline{2-3}
    & control & (singular) foliations  \\
  \hline
\multirow{4}{*}{ 
    \begin{minipage}{0.23\linewidth} \begin{center} \vspace{-1em}
   modern  \\ classical \\ mechanics \\ (PDE) \end{center} \vspace{-2em}
  \end{minipage} }    
& conservation of energy	&    \textbf{multisymplectic}  \\ \cline{2-3}
    & symmetries &  \textbf{Cartan moving frames}\\ \cline{2-3}
    & dissipation / interaction   & Stokes--Dirac \\ \cline{2-3}
    & control & foliations \\
  \hline  
   \end{tabular}
\end{center}
Those structures that are written \textbf{in bold} have discrete analogues, i.e. the numerical methods preserving them can be constructed. First we will walk the reader through the upper part of this table, devoted to finite dimensional mechanical systems, and give some hints on what can be done for the lower part, i.e. for continuous media. 
Then we will present another even more general geometric formalism -- the one of \emph{graded manifolds} -- and explain (or at least try to convince the reader) that it generalizes all the notions simultaneously. We thus believe that finding the discrete analogue of graded geometry is a fruitful direction that potentially permits to produce \emph{geometric integrators} for generic mechanical systems.  

\section{State of the art revisited} \label{sec:geomec}

In this section we will consider several cases of geometric formulations of mechanical problems. We will follow the pattern 
\begin{center}
 \framebox{physics of the system} $\to$ \framebox{geometry} $\to$ \framebox{numerical methods}
\end{center}  at least where all of the steps are known.

\subsection{Conservation laws and symmetries}

Historically, the first instance of so-called \emph{geometric integrators} is related to symplectic structures. 
Consider a manifold $M$ equipped with a symplectic form\footnote{See appendix \ref{sec:geom} for details and definitions of the geometric objects.} $\omega$ and a Hamiltonian vector field $v$ with a Hamiltonian $H$: $\iota_v \omega = d H$, and study the evolution equations governed by $v$.
Because of non-degeneracy of $\omega$, $H$ defines $v$ uniquely, one easily checks that preservation of $\omega$ by the flow of it is equivalent to conservation of the total energy $H$ of the system. A straightforward idea is then to keep track of both in numerical simulations, i.e. produce a discrete vertion of the equations, that ``respects'' the geometric internal structure. Such methods are naturally called \emph{symplectic integrators} (\cite{verlet, yoshida}) and are widely used in for example molecular simulations, where conservation of energy is extremely important. While qualitatively the idea ``preservation of the symplectic forms guarantees the energy condervation'' is true, the precise statement would be that what is conserved is a discrete analogue $H_d$ of $H$, which is eventually different for various symplectic methods. However the difference $(H_d - H)$ can be estimated, and by a proper choice of timestep kept bounded by a given parameter for exponentially long time (\cite{RH}).

It is well known that for finite dimensional conservative mechanical systems the symplectic/Hamiltonian and the Lagrangian descriptions are equivalent (at least under some technical non-degeneracy assumptions). Naturally symplectic integrators have analogues in the Lagrangian description -- they are called \emph{variational integrators} (\cite{MaWe, Gery}). The idea there is to produce a discrete version $L_d$ of the Lagrangian $L$ used in the variational principal. Analyzing the extrema of it, one obtains the discrete version of Euler--Lagrange equations. And in order to recover (whenever possible) the symplectic picture, one needs to construct a discrete version of the Legendre transform.

Those two approaches are now state of the art for at least a couple of decades, but already here we can formulate an open question. The Hamiltonian formalism is defined for Poisson structures as well: given a Poisson bracket $\{\cdot, \cdot\}$ on functions on $M$, or  a Poisson bivector field $\pi$,  the Hamiltonian flow of $H$ is given by $v = \{H, \cdot\}$, or equivalently by the condition $v = \pi^{\sharp} dH$. To the best of our knowledge, neither the Lagrangian description nor the appropriate geometric integrator are known for this setting in the general case; some partial results are given in \cite{KLRS}.

Going further in the formalism of theoretical mechanics one inevitably comes across the study of conserved quantities and integrability. Here again, given a first integral even in the symplectic/Hamiltonian case there is no general method of preserving (a discrete version of) it in the numerical computation. There is however one important situation which does work: the conserved quantities obtained from the Noether's theorem, i.e. associated to the symmetries of the system. Mathematically, symmetries are described by a group acting on the phase space of the system leaving invariant the equations of motion of it, or maybe better to say the space of solutions of them. 
To keep track of them numerically, one considers the infenitesimal version of the theorem, i.e. instead of the Lie group action the symmetries are parametrized by the generators of its Lie algebra. This permits to use the approach of Cartan's moving frames (\cite{olver}): the flow of the system is in a sense equivariant with respect to the action of the  vector fields describing these symmetries. It has been shown (for example in \cite{CRH}) that the numerical methods preserving the symmetries are more robust than those that neglect them. 
Let us note here, that up to non-degeneracy assumptions the Noether's theorem works both ways\footnote{... and there is a huge confusion about that in literature}, i.e. from a conserved quantity one should be able to construct back a generating (generalized) symmetry. So, in principal the method should be applicable to a very general situation of conservation laws, however this ``back'' direction is very rarely explicit.

\subsection{Dirac structures}

This short section is to define the main geometric construction used further for port-Hamiltonian and implicit Lagrangian systems -- the Dirac structure. Since it goes beyond the framework of classical differential geometry and rather fits into what is called generalized geometry or higher structures, we are putting this section to the main text. But as mentioned before, for specialists this is merely to fix the notations, so may be safely skipped. 

Let us consider the so-called Pontryagin bundle $E  = TM \oplus T^*M$ -- the direct sum of the tangent a cotangent bundles (we have the definition in the appendix \ref{sec:geom}). 
On pairs of its sections, i.e. on two couples vector field -- one form,  one defines two natural operations: \\
1) symmetric pairing defined at each point of $M$: 
\begin{equation} \label{pairing}
 <v \oplus \eta, v' \oplus \eta'> = \iota_{v'}\eta + \iota_v\eta'  
 \end{equation}
2) Courant--Dorfman bracket: \\[-1em]
\begin{equation} \label{bracket}
   [v \oplus \eta, v' \oplus \eta'] =  
   [v,v']_{\text{Lie}} \oplus ({\cal L}_v \eta' - \iota_{v'} \rd \eta ).
\end{equation}

 An \emph{almost Dirac structure} ${\mathbb D}$ is a maximally isotropic (Lagrangian) subbundle  ${\mathbb D}$
  of $E$, i.e. a subbundle of $E$ on which the pairing (\ref{pairing}) vanishes identically, and which is of maximal rank equal to $\dim(M)$.
  If moreover the subbundle $\Gamma({\mathbb D})$ is closed with respect to the bracket (\ref{bracket}), it is called a \emph{Dirac structure}.

It is easy to produce a trivial example of a Dirac structure:
   ${\mathbb D} = TM$: the vector field part is governed by the commutator bracket, while nothing happens on the 1-form part. 
Some more interesting examples include a graph of differential 2-form or of a bivector; thus Dirac geometry describes   uniformly symplectic and Poisson manifolds. It is also worth mentioning the examples coming from distributions on $M$ -- those will be important for systems with constraints. 

The first condition related to (\ref{pairing}) is basically studying the linear algebra of the fibers over each point of $M$. The second one  
is more involved, and is sometimes called the integrability condition for ${\mathbb D}$ -- we will comment on it as well.

As a historical remark, let us mention that the Dirac structures were introduced by T.~Courant (\cite{courant}) in his PhD thesis with some initial motivation coming from mechanics: roughly speaking this may be a way to treat simultaneously velocities and momenta of a system, which are obviously related. It turned out however, that a more appropriate description uses double bundles (see the works of W.~Tulczyjew, e.g. \cite{tul}), that is a Dirac structure is constructed on a manifold, which is a bundle itself.

\subsection{Constraints}
Now having introduced the geometric formalism of Dirac structures we can profit from it in several different ways. Recall the construction of variation integrators from above: in \cite{MaWe}, the authors claim that the formalism can be extended to systems with constraints, i.e. to mechanical systems with conditions on coordinates and velocities:
\begin{equation}  \label{eq:constr}
  \varphi^a(q, \dot q) = 0, a = 1, \ldots, m
\end{equation}
This idea has been made explicit in \cite{YoMa1}, and we have considered some examples and improvements in \cite{SH-zamm, RSHD}, all based on Dirac structures.

Let us mention here that geometrically the constraints are described by a distribution $\Delta_Q \subset TQ$ -- i.e. a set of subspaces of the tangent spaces to the configuration space of the system. 
One can actually make an elegant link between holonomic constraints and the integrability of the corresponding Dirac structure.
In this paper we will only illustrate the use of this approach on a scholar example of non-holonomic constraints -- the Chaplygin sleigh leaving the general discussion with proper mathematical details to \cite{KLRS}. We just mention here that the phenomenon of non-holonomic constraints that we observe is an example of the existence of a contact structure, and may open a potentially rich direction in the context of jet spaces (\cite{roubtsov}).

The Chaplygin sleigh is a mechanical system on the plane that is allowed to move only in a direction given by its edge. The configuration space is then given by the position of its contact point $(x, y)$ and the angle $\theta$ with respect to a fixed axis. In these coordinates the Lagrangian reads
$$
   L=\frac{m}{2}\left(\dot{x}^2+\dot{y}^2+\left(\frac{I}{m}+a^2\right)\dot{\theta}^2-2a\sin\theta\dot{x}\dot{\theta}+2a\cos\theta\dot{y}\dot{\theta}\right),
$$
where $m$ is the mass of the sleigh, and $a$ is a geometric parameter representing the distance from the contact point to the center of mass. 
The condition of absence of orthogonal sliding translates into the constraint
$$
  \dot{x}\sin(\theta) - \dot{y}cos(\theta) = 0.
$$
In the Dirac language, this means that the dynamics is governed by $L$ and the constraint distribution $\Delta_Q = Ker(\sin \theta \rd x - \cos\theta \rd y)$. We thus have all the necessary ingredients to apply the Dirac structure based numerical methods. 

\begin{figure}[htp]
\begin{subfigure}{1.\linewidth}
\includegraphics[width=0.55\linewidth]{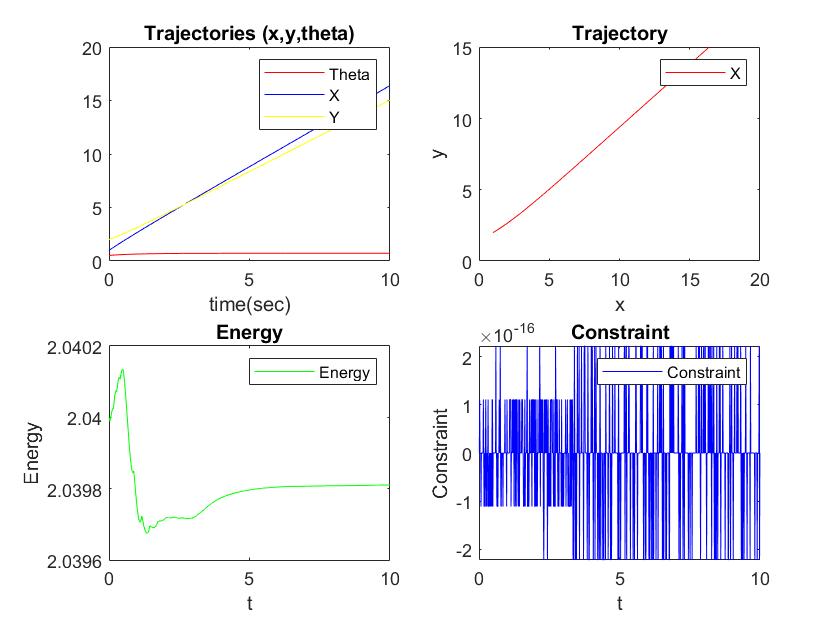}
\caption{The reference solution, constraints satisfied automatically}
\end{subfigure}
\begin{subfigure}{0.49\linewidth}
\includegraphics[width=1.\linewidth]{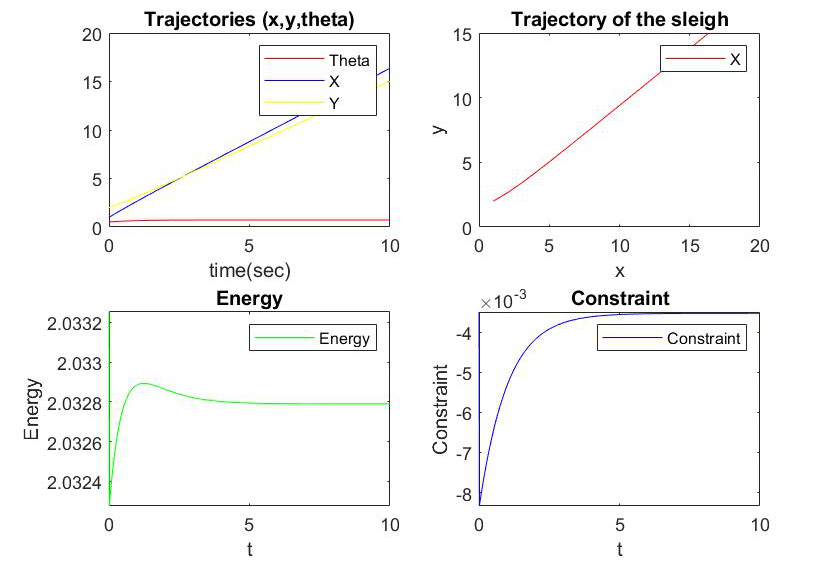} 
\caption{explicit Euler}
\end{subfigure}
\begin{subfigure}{0.49\linewidth}
\includegraphics[width=1.\linewidth]{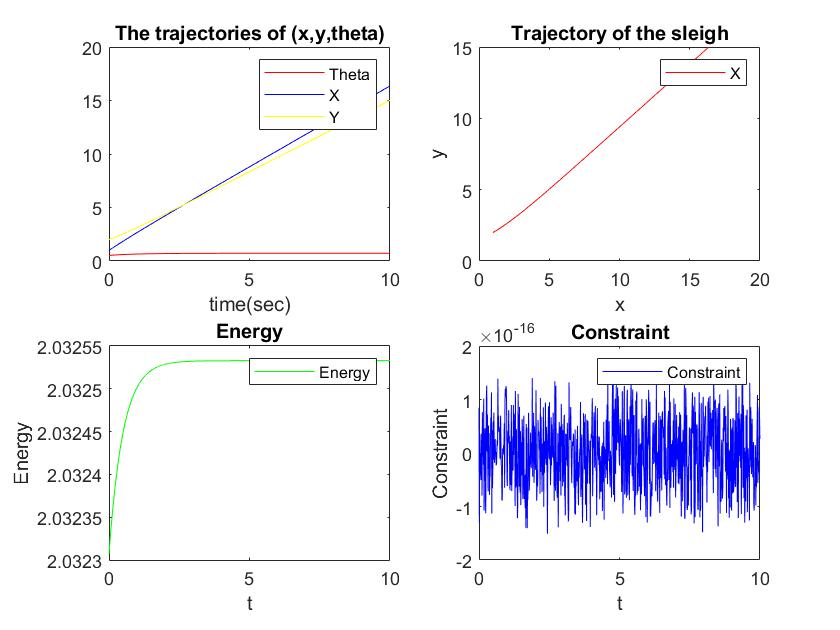}
\caption{Dirac-1}
\end{subfigure}
\caption{Versions of Euler method compared to Dirac-1 from \cite{RSHD}}
\label{fig:plots}
\end{figure}

For this simple system one can actually choose the description in terms of the value of the velocity of the sleigh $v = \sqrt{v_x^2 + v_y^2}$ and compute the components $v_x$ and $v_y$ afterwards, satisfying thus the constraint automatically. We will use this solution as a reference one to compare the accuracy of numerical methods. The figures (\ref{fig:plots}.a -- \ref{fig:plots}.c) and the table \ref{tab:error} clearly show that the Dirac structure based integrators produce more satisfactory results at comparable resources, while for instance the simplecticity is no longer relevant.

\begin{table}[htp]
    \begin{tabular}{ | l | l | l | l | l | l |}
    \hline Method & Error $x$ & Error $y$ & Error $\theta$& Error in constraint& Error in energy  \\ \hline
  explicit Euler & $0.00024 $ & $0.00017$& $ 4.1692\cdot 10^{-6}$& $\sim  10^{-3}$& $ \sim 5\cdot 10^{-5}$ \\ \hline
    symplectic Euler & 0.00003  & 0.00011 &$6.3841 \cdot 10^{-6}$& $\sim  10^{-3}$& $\sim 5\cdot 10^{-5}$  \\ \hline
    Dirac-1 method &  $ 0.00025$ & $ 0.0002$  &  $9.2541 \cdot 10^{-8}$ & $ \sim  10^{-16}$  &  $\sim 5\cdot 10^{-5}$\\ \hline
    \end{tabular}

\caption{Versions of Euler method compared to Dirac-1 from \cite{RSHD}}
\label{tab:error}
\end{table}
Here we obviously consider rather basic methods. A thorough comparison of more advanced (higher order) Dirac structure based numerical methods with the real-life tools will be provided in \cite{LS}.

Let us also mention that the dynamics defined in the original article \cite{YoMa1} by the so-called partial vector field can be made intrinsic using the language of Lie algebroids (\cite{grabowska}), and those  are important examples of differential graded manifold we discuss in section (\ref{sec:graded}).

\subsection{Interaction}

In all the cases discussed above the systems were isolated, it is though very natural to consider interacting systems or introduce some external forces. 
One possible approach to treating those is called \emph{port-Hamiltonian} (\cite{maschke, vdS}).
To the Hamiltonian equations $v = \pi^{\sharp} dH$ one adds all possible internal and external forces obtaining thus a system of the following form
\begin{equation}
  \dot {\mathbf{x}} = (J(\mathbf{x}) - R(\mathbf{x}))\frac{\partial H}{\partial \mathbf{x}} + g(\mathbf{x})\mathbf{f}, \label{eq:port-ham}
\end{equation}
where  $J(\mathbf{x})$ -- an antisymmetic matrix (morally reproducing $\pi$), while  $R(\mathbf{x})$ and $\mathbf{f}$ are new terms responsible for interaction and dissipation. It is important to understand that this approach is not really a new formalism generalizing Hamiltonian systems, it is rather a way to ``order'' the system and  \cite{maschke, vdS} give a classification of the terms one can add together with their physical meaning.  This permits to decompose the system into simple blocks, interacting by transparent rules (via \emph{ports}), which in turn is useful for implementing the simulation algorithms  (see e.g. \cite{falaize}).

The reason why we are interested in this approach is that apparently (\cite{vdS}) port-Hamiltonian systems may also be described using Dirac geometry, namely the defining property of (almost) Dirac structures serves an avatar of power balance. It is in general not known how to preserve an arbitrary Dirac structure in a discretization, however for a large class of port-Hamiltonian systems the structure is given by a graph of a skew-symmetric operator, so is rather similar to the Poisson bivector case. For those we will suggest a promising technique in section (\ref{sec:graded}).

\subsection{Control}
We conclude the overview of finite dimensional systems by briefly mentioning a work in progress related to geometic aspects of control theory. A control system is a system of differential equations of the form 
$$
\dot {\mathbf{x}} = f(\mathbf{x}, \mathbf{u}), 
$$
where $\mathbf{u}$ are control inputs that can be chosen freely. The case of linear controls is rather well studied and gives rise to the theory of \emph{foliations}. We expect that the theory of singular foliations can be of use for non-linear case as well. 

Not going into actual definitions, to give a rough\footnote{We have heard this beautiful explanation from Camille Laurent-Gengoux.} idea, a \emph{regular foliation} may be though of as a lasagna: a decomposition of a 3-dimensional space into 2-dimensional layers (\emph{leaves}). A \emph{singular foliation}, is when into this lasagna someone has dropped a spaghetti or a grain of rice - those will be \emph{singular leaves}. 

For control systems the first natural question to ask is whether every couple of points in the phase space can be connected by an appropriate choice of controls $\mathbf{u}$. If so, the system is called controllable, and the associated foliation is trivial containing only one leaf. Otherwise the phase space decomposes into attainability leaves. The idea (\cite{LRS}) is then to study transversal directions to a singular leaf to figure out what to add to make a system controllable.

\subsection{And what for continuous media?}
The above constructions, as mentioned, concern finite dimensional mechanical systems, contemporary continuous media mechanics however relies heavily on partial differential equations, the phase space of which is essentially infinite dimensional. The results concerning our preferred pattern of geometrization and discretization are only partial there. 

For conservative systems the analogue of Hamiltonian formalism is related to \emph{multisymplectic} geometry, and the appropriate numerical methods do exist. They are mentioned for example in \cite{MaWe}, although not many thorough benchmarking results can be found. Moreover spelling out the multisymplectic form and the n-Hamiltonians of a given system is less straightforward than in the finite dimensional case. 

For the symmetries, however, the machinery is rather well developped and is still based on Cartan's moving frames. The same robustness result as before is typically observed (\cite{CRH}) for symmetry preserving integrators for PDEs.

For other more complicated systems, like the ones with interaction, the results are rather case-by-case. Two important classes are worth mentioning though. The first one is an attempt to apply the port-Hamiltonian logic, where the interaction is done through the common boundary of the domains in the continuous media. This produces the so-called Stokes--Dirac structures, which is however a slight abuse of notations, since the proper infinite dimensional analogue of Dirac structures is not defined.
The second one is when the equations permit a geometric formulation from the very beginning: in terms of differential forms and (co)boundary operators. This results in \emph{Discrete Exterior Calculus} (DEC, \cite{DEC}), which is a state of the art tool in many important situations (see \cite{DEC-aziz} and references therein).

\section{Graded world} 
\label{sec:graded}

In this section we will describe the geometric construction which is even more general than Dirac structures above -- \emph{$Q$-structures} on \emph{graded manifolds}. The motivation to do that is twofold: first -- it potentially permits to treat all the finite dimensional cases from section (\ref{sec:geomec}) in a uniform way, second -- in contrast to most of the above constructions it does have a well defined infinite dimensional analogue. 
We will still stick to the local description, morally replacing manifolds by vector spaces, although in the graded setting this simplification may produce interesting consequences (\cite{KS-graded}).

\subsection{Idea of the definition}
One says that a manifold $M$ is \emph{graded} if it is equipped with a \emph{grading}, i.e. to each coordinate  $x^i$ on $M$ one can assign a label $deg(x^i) \in \mathbb{Z}$, called its \emph{degree}, and this can be done globally in a consistent way. For the purpose of this paper it is enough to assume  $deg(x^i)$ to be non-negative for all the coordinates. The general situation is more involved since the functional space becomes more complicated (see appendix in \cite{DGLG} and \cite{KS-graded}).

For the moment this grading can be treated formally, and it is only responsible for defining the commutation relations between the coordinates. In contrast to the classical case, 
$$x^i \cdot x^j = (-1)^{deg(x^i)deg(x^j)} x^j \cdot x^i.$$ 
Moreover the gradings are compatible with algebraic operations:
$$deg(x^i\cdot x^j) = deg(x^i) + deg(x^j).$$ 
The most general possible function on such a graded manifold would be a formal power series depending on all the variables. But for interesting cases it is enough to consider polynomials in non-zero degree variables with smooth coefficients depending on zero-degree ones. For them the notion of homogeneous functions is well-defined, which produces the commutation relations:
$$f \cdot g = (-1)^{deg(f)deg(g)}g \cdot f.$$

A lot of objects and operations from classical differential geometry can be defined on graded manifolds ``out of the box'', meaning that they work almost as usual with the only essential difference: every time when graded quantities are being permuted a sign may appear non-trivially depending on their degrees. 
For example a graded vector field $v$ of even degree $deg(v)$ is automatically self-commuting:
$$[v, v] \equiv vv - (-1)^{deg(v)deg(v)} vv = 0.$$ 
Meanwhile self-commuting of a vector field $Q$ of an odd degree produces a non-trivial condition: 
$$[Q, Q] \equiv QQ - (-1)^{deg(Q)deg(Q)} QQ = 2 Q^2.$$ 
On a graded manifold, a self-commuting vector field of degree $1$ is called a \emph{$Q$-structure}. 
A graded manifold equipped with such a vector field is called a \emph{differential graded manifold}, or a  \emph{$Q$-manifold}, the latter name often appears in physics literature.

Below we describe two classical examples of differential graded manifolds.
 
\subsection{Differential forms}

Consider a tangent bundle to a smooth manifold $\Sigma$, and declare the fiber linear coordinates  $\theta$  to be of degree $1$ -- the usual notation for such a \emph{shifted} tangent bundle is  $T[1]\Sigma$. 
Since $deg(\sigma^{\mu}) = 0$, they commute: $\sigma^{\mu_1}\sigma^{\mu_2} = \sigma^{\mu_2}\sigma^{\mu_1}$ and the degree of any function of them is zero:
 $deg(h(\sigma^{1}, \dots, \sigma^{d})) = 0$. 
 According to our assumption $deg(\theta^{\mu}) = {1}$, i.e. $\theta$'s anticommute:   $\theta^{\mu_1} \theta^{\mu_2} = - \theta^{\mu_2} \theta^{\mu_1}$. A generic homogeneous degree $p$ function  on $T[1]\Sigma$ reads
$$f = \sum f_{\mu_1\dots \mu_p}(\sigma^1, \dots, \sigma^d) \theta^{\mu_1}\dots\theta^{\mu_p}.$$ 
And as above, 
$$f \cdot g = (-1)^{deg(f)deg(g)}g \cdot f.$$ 
The reader has certainly noticed the resemblance of this construction with the one of differential forms from section (\ref{sec:geom}):
 $$f \leftrightarrow \alpha = \sum f_{\mu_1\dots \mu_p} \rd \sigma^{\mu_1} \wedge\dots \wedge\rd \sigma^{\mu_p} \in \Omega(\Sigma).$$

Consider now a vector field   $Q = \sum  \theta^{\mu}\frac{\partial}{\partial \sigma^\mu}$ defined on $T[1]\Sigma$. It is easy to see that   $\deg{Q} = {1}$ and $[Q, Q] \equiv 2 Q^2 = 0$, hence it is a $Q$-structure, mimicking exactly the properties of the De Rham differential. This explains the name of \emph{differential} graded manifolds. This also provides an alternative (certainly, not the easiest) way to define differential forms from scratch.

\subsection{Poisson bivector}
Consider a Poisson manifold $M$ (see the appendix (\ref{sec:geom}) to recall the notations), and consider now the shifted cotangent bundle $T^*[1]M$ -- the cotangent bundle to $M$ again with fiber linear coordinates of degree $1$, i.e. $deg(x^i) = 0$, $deg(p_i) = 1$. It carries a canonical (graded) symplectic form $\omega = \sum_i dp_i \wedge dx^i$. Recall that the Poisson structure on $M$ can be encoded in a bivector field $\sum \pi^{ij}\partial_i \wedge \partial_j$, which in turns defines a function on  $T^*[1]M$: $H = \frac12\pi^{ij}p_ip_j$. The (graded) Hamiltonian vector field corresponding to this $H$ reads
\begin{equation} \label{Qpi}
   Q_{\pi} = % \left\{\frac12\pi^{ij}p_ip_j , \cdot \right \}_{T^*M} = 
   \pi^{ij}(x)p_{j} \frac{\partial}{\partial x^i} - 
      \frac{1}{2} \frac{\partial \pi^{jk}}{\partial x^i}p_{j}p_{k} \frac{\partial}{\partial p_i}
\end{equation}
It is instructive to check that  $deg(Q_{\pi}) = 1$, and that the condition $Q_{\pi}^2 = 0$ is equivalent ot the Jacobi identity (\ref{jacobi}).

\subsection{Potential applications}

The two example of differential graded manifolds given above look specific, but they are actually rather generic, and in particular here we can already explain how this can be useful for mechanics. So, this section is here to put together all the pieces of the puzzle. 

It is now clear that graded geometry provides a uniform description of rather different objects such as differential forms and Poisson geometry. So, going through the table in the introduction we see that we obtain a description of symplectic and Poisson structures ``for free''. 

An attentive reader might already have noticed that the construction goes beyond. Namely for port-Hamiltonian systems the almost Dirac structure in the game is (often) a graph of a skew-symmetric operator (i.e. a bivector). Thus one can always consider an appropriate $T^*[1]M$ (maybe with a different $M$) as a graded manifold, and produce a degree $1$ vector field on it as above -- equation (\ref{Qpi}), by some language abuse we call it \emph{almost $Q$-structure}. This vector field squaring to zero will amount to the integrability condition, i.e. the almost Dirac structure being actually Dirac.

In fact for any Dirac structure one can produce a natural graded manifold equipped with a $Q$-structure. We will not describe this construction here, referring to a general statement in  \cite{KSS}. Note just that for an almost Dirac structure one again obtains an almost $Q$-structure, covering thus in particular the case of generic constraints. This is an honest $Q$ structure when the structure is actually Dirac, that is constraints are holonomic. Moreover, one can associate $Q$-structures to singular foliations as well (\cite{LLS}), hence covering the control theory problems. 

The previous paragraph basically means that the internal geometry of systems from all the upper part of the table from the introduction can be naturally encoded using the language of differential graded manifolds -- this is the main conceptual message of this paper. It is obviously legitimate to ask how this can be useful for designing numerical methods preserving this internal geometry. Two important remarks are in place here.

\begin{figure}[htp]
\includegraphics[width=0.8\linewidth, trim=20 380 20 0, clip]{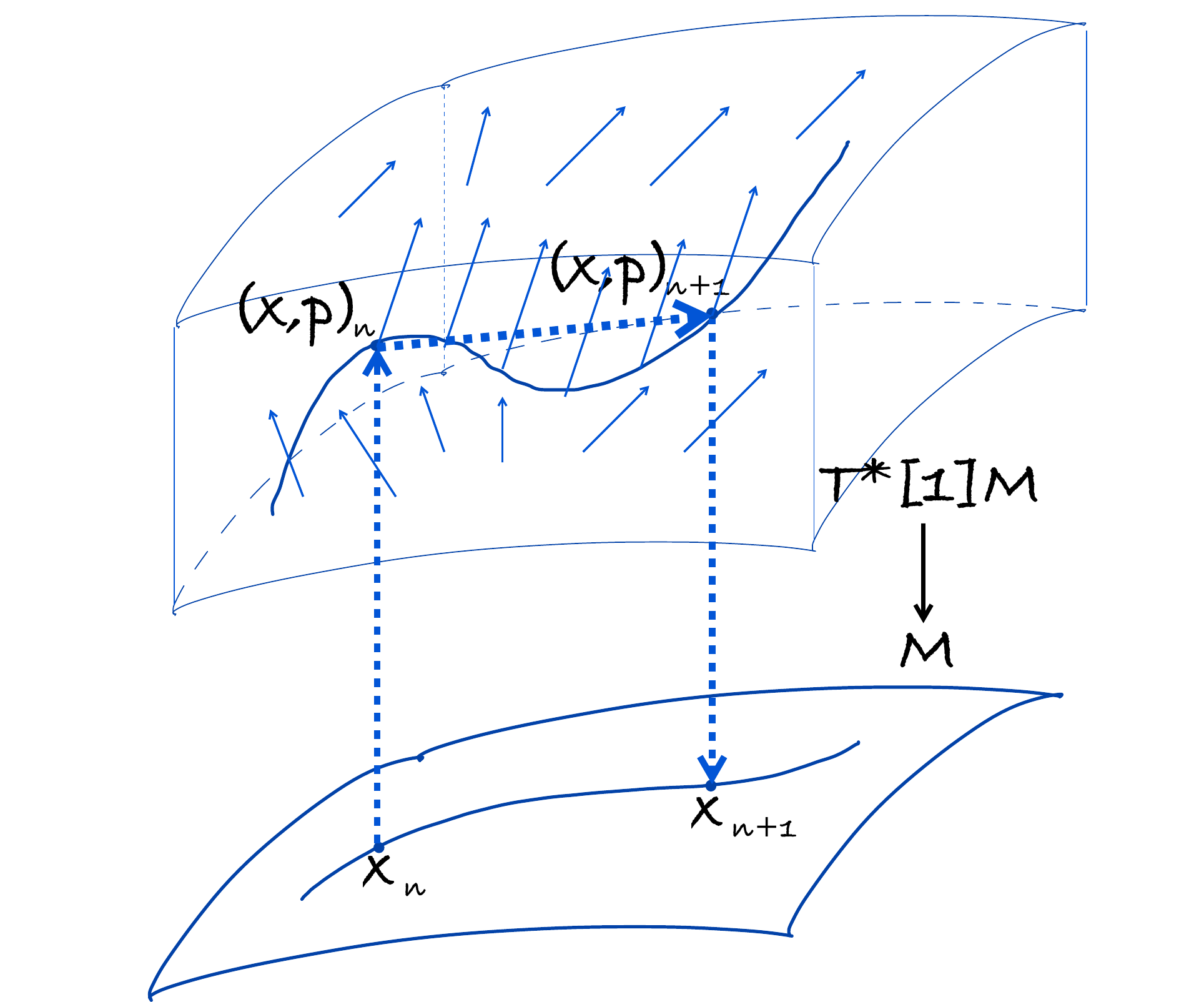}
  \caption{Constructing a $Q$-structure preserving integrator --  a schematic representation.}
    \label{fig:lift}
\end{figure}

First, recall the Cartan's moving frame construction: the vector fields generating the symmetries were ``interacting'' with the vector field generating the dynamics. And recall that an almost $Q$-structure is merely a vector field, in the generic case on the shifted cotangent bundle $T^*[1]M$ to the (eventually extended) phase space $M$ of the mechanical system. There is a canonical way to lift the dynamics from $M$ to its cotangent bundle, using the Lie derivative along the vector field generating it. Hence the discretization procedure will look roughly like this (see Fig. \ref{fig:lift}):
\begin{center}
\framebox{start from an initial point on $M$, lift it to $T^*[1]M$} $\to$ \\ 
$\to$ \framebox{spell-out the commutation relation of the lift of the dynamics with a $Q$-structure} $\to$ \\$\to$ \framebox{produce the next point of the discrete flow on $T^*[1]M$} $\to$ \\$\to$ 
\framebox{project it back to $M$}
\end{center}

Second, recall that an honest $Q$-structure, i.e. a vector field squaring to zero, resembles the De Rham differential. And for it the Discrete Exterior Calculus is well developped. Inspired from DEC, the natural idea is then to profit from the similar constructions of equivariant $Q$-cohomology (\cite{KS, VS-jgp}) in order to take into account the squaring to zero condition where it is appropriate. 

These two ideas are the main technical messages of this paper, they will be detailed soon in \cite{SH-graded}. Last but not least, let us very briefly repeat that in contrast to most of the above classical geometric constructions, the $Q$-structures are also well defined on infinite dimensional graded manifolds. For example, the famous article \cite{aksz} describes precisely the way to define the structure of a differential graded manifold on a space of mappings between differential graded manifolds. Such tools are widely used in high energy physics, and we expect the application to mechanics to be a regression of those. 

\section{Conclusion}
In this paper we have given an overview of geometric structures relevant for qualitative analysis of mechanical systems, including rather advanced ``discoveries'' from generalized and graded geometry. We hope that the level of details permits if not to start applying those, but at least to be convinced that they are not completely artificial. An interested reader is invited to follow the given references or even consider this text as a call for collaboration, since some of the described problems are open or at least not explored enough.

\vspace{1em}
\textbf{Acknowledgments.} \\
This text is inspired by a lot of conversations during several CITV -- Souriau Colloquium meetings, V.S. and A.H. would like to thank all the participants of them for creating this working atmosphere. V.S. personally thanks Wlodzimierz Tulczyjew for enlightening discussions at various stages of this work. Some parts of this research and collaborations have been supported by the career starting grants from LaSIE (2019) and INSIS CNRS (2018), as well as the Young Researcher Grant from La Rochelle University (ACI project call 2018 and 2019). The mentioned ongoing work is supported by the CNRS 80Prime GraNum project.

\appendix

\section{Differential geometry -- minimal working knowledge}
\label{sec:geom}

This appendix recapitulates the minimal working knowledge from differential geometry needed to understand the constructions of the article. A trained geometer can obviously skip the whole section, for all the others we provide a concise description, eventually skipping or simplifying some details. In particular we will use the notions of manifolds and bundles without properly defining them -- to get an idea one should just think (locally) of vector spaces of appropriate dimension.

\subsection{Vector fields} 

Thus let $M$ be a \emph{manifold} of dimension $n$, denote $TM$ -- its \emph{tangent bundle} (i.e. all the tangent vectors to it at each point) and $T^*M$ -- its \emph{cotangent bundle}. 
Locally $TM$ can be viewed as  
$$
TM \simeq \mathbb{R}^{2n} = \mathbb{R}^n \times V,
$$ 
where $V$ is a vector space of the same dimension as  $M$, then 
$$
T^*M \simeq \mathbb{R}^{2n} = \mathbb{R}^n \times V.^*
$$
The set of \emph{sections} of $TM$ is denoted by $\Gamma(TM)$ -- those are vector fields on $M$. One can think for example of a velocity field on a surface.   On these sections a binary operation is defined: their \emph{commutator} 
$$[\cdot, \cdot] \; \colon \; \Gamma(TM)\times \Gamma(TM) \to \Gamma(TM).$$ 
In components it reads:
$$[v, w]^i = \sum_j (v^j \partial_j w^i - w^j \partial_j v^i),$$
 where $\partial_j$ --  is the derivative by the $j$-th coordinate. For the mechanical intuition, given two velocity fields, one can follow the flow of one of them, then the other one, or in the reversed order, the result is typically not the same --  the commutator quantifies this difference infinitesimally.

\subsection{Differential forms / multivector fields} 

On a vector space $V$ one defines skew-symmetric $k$-linear forms. When this is done at each point of $M$, under some regularity assumptions, one obtains the object called a \emph{differential $k$-form}, or a \emph{differential form of degree $k$}, i.e. a skew-symmetric ``function'', the ``arguments'' of which are  $k$ vector fields on  $M$. 
Usual functions on $M$ can be viewed as  $0$-forms. The standard examples of $1$-forms, i.e. covector fields, are the objects dual to vector fields $\partial_i$, they are denoted  $dx^i$. The duality is understood in the sense that  $dx^i(\partial_j) = \delta^i_j$, where $\delta^i_j$ -- is the Kronecker's symbol.
In dimension $2$ an example of a $2$-form would be the oriented area.

Two natural operations are defined for differential form: 
\begin{itemize}
  \item \emph{Contraction} of a vector field $v$ with a form $\alpha(\cdot, \cdot, \dots)$ : 
  $$
  \iota_v(\alpha) := \alpha(v, \cdot, \dots).$$ 
   It lowers the form degree. 
  \item \emph{Exterior (De Rham) differential} of a form $\alpha$: 
 \begin{eqnarray} \nonumber
  &&d \alpha (v_0, v_1, \dots, v_k) :=   \\
&& \sum_i (-1)^i v_i \alpha(v_0,  \dots, \hat{v_i} \dots, v_k) + 
 \sum_{i<j} (-1)^{i+j} \alpha([v_i, v_j], v_0,  \dots, \hat{v_i}, \dots \hat{v_j}, \dots, v_k),   \nonumber
 \end{eqnarray}
  where $\hat{v}$ denotes omission of the argument. The differential raises the form degree.
   \end{itemize}

The notation $dx^i$ above is not an abuse: the objects dual to  $\partial_i$ are actually differentials of coordinate functions.
 If $d \alpha \equiv 0$, the form  $\alpha$ is called \emph{closed}, and when  
   $\alpha$ is itself a differential of some other form $\beta$: $\alpha = d\beta$, then  $\alpha$ is called  \emph{exact}, and the form $\beta$ is sometimes called its  \emph{integral}.
   An interesting property of the De Rham differential is that it squares to zero: $d d \beta \equiv 0$ for any from $\beta$. This means that any exact form is closed, but not necessarily vice versa. 

If a $2$-form is non-degenerate (at each point, in the linear algebra sense), it is called  \emph{almost symplectic}; and if moreover it is closed, then \emph{symplectic}. A vector field  $v$, the contraction of which with a symplectic form  $\omega$ is exact with an integral  $H$ ($\iota_v \omega = d H$), is called  a \emph{Hamiltonian vector field with the Hamiltonian function $H$}.

Note that a similar (or better to say dual) construction is possible on a cotangent bundle -- the result would be multivector fields. And the analogue (or in a sense the inverse) of a symplectic form would be the Poisson bivector. 

 Let a manifold  $M$ be equipped with a Poisson bracket, i.e. a skew-symmetric operation on the space of functions on $M$:
  $\{\cdot, \cdot \} \colon C^{\infty} (M) \times C^{\infty} (M) \to C^{\infty} (M)$, satisfying the Leibniz property:
 $$
 \{f, gh\} = \{f,g\}h + g\{f, h\}
 $$
and the Jacobi identity
 $$
\{f,\{g, h\}\} + \{g,\{h, f\}\} + \{h,\{f, g\}\} = 0
$$
  
From the geometric point of view, the Poisson bracket can be rewritten as 
$\{f, g \} = \pi(\rd f, \rd g)$, where $\pi \in \Gamma(\Lambda^2 TM)$ -- a bivector field with components  $\pi^{ij}(x) = \{x^i, x^j\}$. A bivector field $\pi^{ij}\partial_i \wedge \partial_j$ is a biderivation, so it satisfies the Leibniz property automatically.The Jacobi identity in components of  $\pi$ reads
\begin{equation}  \label{jacobi}
\sum_l \left(\displaystyle \frac{\partial \pi^{ij}(x)}{\partial x^l}\pi^{lk}(x) 
+ \displaystyle \frac{\partial \pi^{ki}(x)}{\partial x^l}\pi^{lj}(x) +  
 \displaystyle \frac{\partial \pi^{jk}(x)}{\partial x^l}\pi^{li}(x) \right) = 0.
\end{equation}
We use this description in the context of graded geometry in section (\ref{sec:graded}).

\end{document}